\documentclass[aps,prd,twocolumn,groupedaddress,showpacs,nofootinbib]{revtex4}
\usepackage{graphicx}

\begin{document}

\title{Inflessence\,: a phenomenological model for inflationary quintessence}

\author{V.F. Cardone}
\thanks{Corresponding author, email: {\tt winny@na.infn.it}}
\author{A. Troisi}
\affiliation{Dipartimento di Fisica ``E.R. Caianiello'', Universit\`{a} di Salerno and INFN, Sez. di Napoli, Gruppo Coll. di Salerno, via S. Allende, 84081 - Baronissi (Salerno), Italy}

\author{S. Capozziello}
\affiliation{Dipartimento di Scienze Fisiche, Universit\`{a} di Napoli ``Federico II'' and INFN, Sez. di Napoli, Compl. Univ. Monte S. Angelo, Edificio N, Via Cinthia, 80126, Napoli, Italy}

\begin{abstract}

We present a phenomenologically motivated model which is able to give rise both to an inflationary epoch and to the present day cosmic acceleration. We introduce an approach where the energy density depends on the scale factor $a$ in such a way that a smooth transition from the inflation to the matter dominated era and then to the nowadays accelerated expansion is achieved. We use the dimensionless coordinate distance $y(z)$ to the Gold Type Ia Supernovae dataset and to a sample comprising 20 radio galaxies, the shift parameter ${\cal{R}}$ and the acoustic peak parameter ${\cal{A}}$ to test whether the model is in agreement with observations and to constrain its main parameters. As an independent cross check, we also compare model predictions with the lookback time to galaxy clusters and the age of the universe. Although phenomenologically inspired, the model may be theoretically motivated either resorting to a scenario with a scalar and phantom fields (eventually interacting) or as an effective description arising from higher order gravity theories.

\end{abstract}


\maketitle

\section{Introduction}

An impressive amount of astrophysical data has been accumulated in recent years opening the era of the {\it observational cosmology} and leading to a complete revision of what is thought about the universe. The Hubble diagram of Type Ia Supernovae (hereafter SNeIa) \cite{SNeIa,Riess04}, the anisotropy spectrum of the cosmic microwave background radiation (hereafter CMBR) \cite{CMBR,WMAP,VSA}, the matter power spectrum determined from the clustering properties of the large scale distribution of galaxies \cite{LSS} and the data on the Ly$\alpha$ emitting regions \cite{Lyalpha} all provide indications that the universe have to be described as a spatially flat manifold where matter and its fluctuations are isotropically distributed and represent only about $30\%$ of the overall content. In order to fill the gap and drive the acceleration, a dominant contribution coming from a homogeneously distributed negative pressure fluid has been invoked, usually referred to as {\it dark energy}.

Notwithstanding the astonishing flood of papers on this topic, it is still unclear what are the nature and the nurture of this mysterious component. The simplest explanation claims for the cosmological constant $\Lambda$ thus leading to the so called $\Lambda$CDM model \cite{Lambda}. Although being the best fit to most of the available astrophysical data \cite{WMAP,Teg03,Sel04}, the $\Lambda$CDM model is also plagued by many problems on different scales. If interpreted as vacuum energy, $\Lambda$ is up to 120 orders of magnitudes smaller than the predicted value. Furthermore, one should also solve the {\it coincidence problem}, i.e. the nearly equivalence of the matter and $\Lambda$ contribution to the total energy density. 

As a response to these problems, much interest has been devoted to models with dynamical vacuum energy, dubbed {\it quintessence} \cite{QuintFirst}. These models typically involve scalar fields with a particular class of potentials, allowing the vacuum energy to become dominant only recently (see \cite{QuintRev} for comprehensive reviews). Although quintessence by a scalar field is the most studied candidate for dark energy, it generally does not avoid {\it ad hoc} fine tuning to solve the coincidence problem.

Relying on the observation that there is actually no clear evidence that dark energy is not the same as dark matter, many models have been proposed in which a single fluid accounts for both of them. Collectively referred to as {\it unified dark energy} (UDE) models, these proposals resort to standard matter with exotic equation of state (hereafter EOS) as the only candidate to both dark matter and dark energy thus automatically solving the coincidence problem. The EOS is then tuned such that the fluid behaves as dark matter at high energy density and quintessence (or $\Lambda$) at the low energy limit. Interesting examples are the Chaplygin gas \cite{Chaplygin}, the tachyonic field \cite{tachyon}, the Hobbit model \cite{Hobbit} and the Van der Waals quintessence \cite{VdW}.

Actually, there is still a different way to face the problem of cosmic acceleration. As  stressed in Lue et al. \cite{LSS03}, it is possible that the observed acceleration is not the manifestation of another ingredient in the cosmic pie, but rather the first signal of a breakdown of our understanding of the laws of gravitation. From this point of view, it is thus tempting to modify the Friedmann equations to see whether it is possible to fit the astrophysical data with a model comprising only the standard matter. Interesting examples of this kind are the Cardassian expansion \cite{Cardassian} and the DGP gravity \cite{DGP}. Moving in this same framework, it is possible to find alternative schemes where a quintessential behavior is obtained as a byproduct of fourth order theories of gravity obtained by replacing the Ricci scalar curvature $R$ in the gravity Lagrangian with a generic function $f(R)$ giving rise to a {\it curvature quintessence} \cite{noicurv,MetricRn,PalRn,lnR,Allemandi}.

Despite their radically different underlying philosophies, the above scenarios share a common feature. They all consider cosmic acceleration as a unique event in the evolutionary history of the universe and thus focus their attention only to a limited redshift range. Actually, another period of accelerated expansion took place in the very early universe, namely during the inflationary epoch. Indeed, it was inflation that has suggested to consider a dynamical scalar field as the candidate to explain the present day cosmic acceleration. An immediate step along this direction consists in wondering whether the scalar field today driving the universe is the same as the one which originated the inflationary expansion. Actually, as scalar fields are not the unique possibility to explain the late time acceleration, in the same way they are not the only candidates to give rise to the inflation with extended theories of gravity as valid alternatives. 

Recently, some attempts have been proposed to explain today accelerating universe as a byproduct of primordial inflation \cite{Riotto} without invoking any dark energy model, but serious objections have been raised \cite{controRiotto} making this approach controversial.

From this short overview, it is clear that there is a wide confusion among rival theories for both on inflation and dark energy. Nevertheless, it is highly desirable to find a mechanism to unify both these fascinating issues under the same physical scenario. In order to solve this problem, we adopt here a phenomenological approach. Rather than proposing a somewhat theoretically motivated model, we introduce a single fluid whose energy density dependence on the scale factor is assumed {\it a priori} in such a way that it gives rise to a smooth transition from the inflationary epoch to the matter dominated era finally leading to the present day accelerated expansion. After having tested the model against observations, we also present two possible physical interpreted in terms of two radically different scenarios.

The plan of the paper is as follows. In Sect.\,II we present the main feature of the class of phenomenological models we propose, while Sect.\,III is devoted to the study of the dynamics of the universe in presence of this fluid. The matching with observations is performed in Sect.\,IV where we consider the dimensionless coordinate distance $y(z)$ to the Gold SNeIa dataset and a sample of 20 radio galaxies, the shift parameter ${\cal{R}}$ (related to the distance to the last scattering surface) and the acoustic peak parameter ${\cal{A}}$. As an independent cross check, we also compare the model predictions with the estimated age of the universe and the lookback time to galaxy clusters. Two possible theoretical scenarios to recover our model as an effective description are presented in Sect.\,V, while we summarize and conclude in Sect.\,VI.

\section{The inflessence model}

Current theoretical schemes and observational evidences point out that the evolutionary history of the universe comprises two periods of accelerated expansion, namely the inflationary epoch and the present day dark energy dominated phase. In both cases, the expansion is usually interpreted as the result of the presence of a negative pressure fluid dominating the energy budget. It is natural wondering whether a single (effective) fluid may indeed be responsable of both periods of accelerated expansion. At the same time, this fluid should be subdominant during the radiation and matter dominated epochs in order to not interfere with baryogenesis and structure formation. While it is quite difficult to theoretically formulate the properties of such a fluid, it is, on the contrary, clear what are its phenomenological features. Motivated by this consideration, we propose here to start from the phenomenology of the fluid in an attempt to recover its fundamental properties. 

Let us thus assume that the energy density as function of the scale factor $a$ reads\,:

\begin{equation}
\rho(a) = {\cal{N}} a^{-3} \left ( 1 + \frac{a_I}{a} \right )^{\beta} 
\left ( 1 + \frac{a}{a_Q} \right )^{\gamma}
\label{eq: rhovsa}
\end{equation}
with ${\cal{N}}$ a normalization constant, $(\beta, \gamma)$ slope parameters and $a_I << a_Q$ two scaling values of the scale factor. For later applications, it is convenient to rewrite Eq.(\ref{eq: rhovsa}) in terms of the redshift $z = 1/a - 1$ (having set $a_0 = 1$ with the subscript $0$ denoting henceforth quantities evaluated at the present day, i.e. $z = 0$)\,:

\begin{equation}
\rho(z) = {\cal{N}} (1 + z)^3 \left ( 1 + \frac{1 + z}{1 + z_I} \right )^{\beta}
\left ( 1 + \frac{1 + z_Q}{1 + z}\right ) \ , 
\label{eq: rhovsz}
\end{equation}
having defined\,:

\begin{equation}
z_I = 1/a_I - 1 \ ,
\label{eq: defzi}
\end{equation} 
\begin{equation}
z_Q = 1/a_Q - 1 \ .
\label{eq: defzq}
\end{equation} 
From Eq.(\ref{eq: rhovsa}), it is quite easy to see that\,:

\begin{equation}
\left \{
\begin{array}{ll}
\rho \sim a^{-(\beta + 3)} & {\rm for} \ \ a << a_I << a_Q \nonumber \\
~ & ~ \nonumber \\
\rho \sim a^{-3} & {\rm for} \ \ a_I << a << a_Q \nonumber \\ 
~ & ~ \nonumber \\
\rho \sim a^{\gamma - 3} & {\rm for} \ \ a_I << a_Q << a \nonumber
\end{array}
\right . \ .
\end{equation}
The energy density of the proposed fluid scales as that of dust matter in the range $a_I << a << a_Q$. As we will see, this means that the fluid follows matter along a large part of the universe history, while it scales differently only during the very beginning ($a << a_I$) and the present period ($a >> a_Q$). Moreover, choosing $\beta = -3$, the fluid energy density remains constant for $a << a_I$ thus behaving as the usual cosmological constant $\Lambda$ during the early epoch of the universe evolution. Finally, the slope parameter $\gamma$ determines how the fluid energy density scales with $a$ in the present epoch. 

It is still more instructive to look at the equation of state $w \equiv p/\rho$ where $p$ is the fluid pressure. To this aim, we first remember the continuity equation\,:

\begin{equation}
\dot{\rho} + 3 H (\rho + p) = 0 \ ,
\label{eq: cont}
\end{equation}
with $H = \dot{a}/a$ the Hubble parameter. Using the obvious relation $d\rho/dt = d\rho/da \times da/dt$, it is immediate to get the following expression for the pressure\,:

\begin{equation}
p = - \frac{1}{3} \left ( a \frac{d\rho}{da} + 3 \rho \right )
\label{eq: eqp}
\end{equation}
which holds whatever is the cosmological model provided that $H$ is not vanishing everywhere (i.e. the universe is not stationary). Actually, if the model contains more than a single fluid, this relation strictly holds only for the total pressure and the total energy density. However, it is still valid for each fluid if each one satisfies the continuity equation separately. This happens whenever the fluids are not interacting as in many quintessence models. It is worth noting, however, that an interaction term between the dark matter and the dark energy component is an efficient tool to solve the cosmic coincidence problem. Nevertheless, as discussed by Zimdhal \cite{Zim}, the decay rate of one substance into the other one enters explicitly only in the second derivative of the Hubble parameter and is thus quite difficult to be dtected. Moreover, it is also possible to show that any cosmic EOS $w$ can be interpreted as an effective EOS of an interacting mixture of dust matter and a dark energy component with a suitably assigned EOS $w_X$. This degeneracy among interacting and non interacting models ensures that assuming that our assumption of non interacting fluids does not lead to any loss of generality. Motivated by this consideration, we may therefore assume that Eq.(\ref{eq: eqp}) holds for each fluid separately and use it to determine the EOS of our proposed component.

\begin{figure}
\centering \resizebox{8.5cm}{!}{\includegraphics{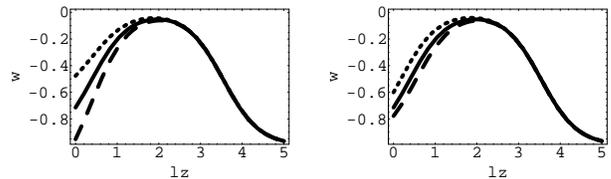}}
\caption{The equation of state $w$ as function of $lz = \log{(1 + z)}$ for models with $(\beta, z_I) = (-3, 3454)$. In the left panel, $z_Q = 1.5$ is set and the three curves refer to $\gamma = 2$ (short dashed), $\gamma = 3$ (solid) and $\gamma = 4$ (long dashed). In the right panel, instead, we set $\gamma = 3$ and use three different values of $z_Q$, namely $z_Q = 0.5, 1.5, 2.5$ (short dashed, solid and long dashed respectively).}
\label{fig: wz}
\end{figure}

Inserting Eq.(\ref{eq: rhovsa}) into Eq.(\ref{eq: eqp}), after some algebra we get\,:

\begin{equation}
w = \frac{\beta}{3} \left ( \frac{1 + z}{2 + z + z_I} \right ) - \frac{\gamma}{3} \left ( \frac{1 + z_Q}{2 + z + z_Q} \right )
\label{eq: wvsz}
\end{equation} 
which is plotted in Fig.\,\ref{fig: wz} as function of $\log{(1 + z)}$ for models with different values of $\gamma$ and $z_Q$ having set $\beta = -3$ and $z_I = 3454$ (see later for the motivations of this choice). It is worth noting that $w$ does not depend neither on $\gamma$ nor on $z_Q$ for high values of $z$ which is expected looking at Eq.(\ref{eq: wvsz}). On the contrary, these two parameters play a key role in determining the behaviour of the EOS over the redshift range $(0, 100)$ which represents most of the history of the universe (in terms of time). 

The role of the different quantities $(\beta, \gamma, z_I, z_Q)$ is better understood considering the asymptotic limits of the EOS. We easily get\,:

\begin{equation}
\lim_{z \rightarrow \pm \infty}{w(z)} = \frac{\beta}{3}
\label{eq: wlim}
\end{equation} 
where $z \rightarrow -\infty$ refers to the asymptotic future. Eq.(\ref{eq: wlim}) shows that setting $\beta = -3$, the fluid EOS asymptotically approaches that of the cosmological constant, i.e. $w_{\Lambda} = -1$. In general, if we impose the constraint $\beta < -1$, we get a fluid having a negative pressure in the far past so that it is able to drive the accelerated expansion occurring during the inflationary epoch. 

It is now clear that $z_I$ controls the transition towards the past asymptotic value in the sense that the larger is $z$ with respect to $z_I$, the smaller is the difference between $w(z)$ and its asymptotic limit $\beta/3$. This consideration suggests that $z_I$ has to take quite high values (indeed, far greater than $10^3$) since, for $z >> z_I$, the universe is in its inflationary phase. 

Let us now consider the present day value of $w$ that turns out to be\,:
\begin{equation}\
w_0 = \frac{\beta}{3 (2 + z_I)} - \frac{\gamma}{3} \left ( \frac{1 + z_Q}{2 + z_Q} \right ) 
\simeq -\frac{\gamma}{3} \left ( \frac{1 + z_Q}{2 + z_Q} \right )
\label{eq: wz}
\end{equation} 
where, in the second line, we have used the fact that $z_I$ is very large. Being $z_Q > 0$, in order to have a present day accelerated expansion, $w_0$ should be negative so that we get the constraint $\gamma > 0$. Moreover, depending on the values of $\gamma$ and $z_Q$, $w_0$ could also be smaller than $w_{\Lambda}$ so that we may recover phantom\,-\,like models \cite{phantom}. The parameter $z_Q$ then regulates the transition to the dark energy\,-\,like dominated period. 

Summarizing, the fluid with energy density and EOS given by Eqs.(\ref{eq: rhovsz}) and (\ref{eq: wvsz}) is able to drive the accelerated expansion of the universe during both the inflationary epoch and the present day period. Therefore, such a fluid play the role of both the {\it inflaton} and the {\it quintessence} scalar field so that we dub it {\it inflessence} (contracting the words {\it inflationary quintessence}).

It is worth noting that, since $\rho$ scales with $a$ as the dust matter energy density for a long period of the universe history, the coincidence problem is partially alleviated. Moreover, although $w_0$ could be smaller than -1 as for a phantom field, the EOS asymptotically tends to a value larger than -1 (provided that $-3 \le \beta \le -1$) so that it is likely that the Big Rip is significantly delayed or does not occur at all.

\section{The dynamics of the universe}

Having worked out the main physical features of the model, we are now able to study the dynamics of the universe. To this aim, we assume spatial flatness in agreement with the CMBR observations \cite{CMBR,WMAP,VSA}. As a first step, we have to to decide what are the ingredients of the cosmic pie. Since we are interested to the full evolutionary history of the universe, we cannot neglect the radiation contribution even if including or neglecting this term will have no impact at all on the cosmological tests we will perform later. An obvious ingredient is represented by the baryons since they make up the structures (galaxies and cluster of galaxies and the hot intracluster gas) we observe today. It is more subtle to decide whether including or not dark matter. As we have seen, the inflessence energy density scales as that of dust matter along the most of the universe history decoupling from it only recently (i.e., for $z << z_Q$). From this point of view, inflessence should be considered as a candidate for a unified dark energy model, similar to the (generalized) Chaplygin gas \cite{Chaplygin} or the Hobbit model \cite{Hobbit}, with the further attracting feature of unifying also inflation with dark matter and dark energy. If this interpretation is correct, however, the inflessence EOS should be almost vanishing over a certain redshift range. Solving $w(z_M) = 0$, we get\,:

\begin{equation}
z_M = \frac{y_Q \gamma - (2 + y_Q) \beta \pm \sqrt{y_Q {\cal{Z}}(\beta, \gamma, z_Q, z_I)}}{2 \beta}
\label{eq: zm}
\end{equation}
with $y_Q = 1 + z_Q$ and

\begin{equation}
{\cal{Z}}(\beta, \gamma, z_Q, z_I) =  y_Q \beta^2 + 2 (2 - y_Q + 2 z_I) \beta \gamma + y_Q \gamma^2 \ .
\label{eq: defzm}
\end{equation}
It is easy to check that, for reasonable values of $\beta$ and $z_I$, $z_M$ is always a complex number and hence the EOS never vanishes and is always negative. Thus, even if its energy density scales as that of dark matter over the most of the universe life, inflessence cannot play the same role of matter since its equation of state is always significantly different from null. As a result, we have to include also the dark matter in the total energy budget.

Introducing now the critical energy density $\rho_{crit} = 3 H_0^2/ 8 \pi G$, the expansion rate $H$ is determined by\,:

\begin{equation}
E^2(z) = \Omega_r (1 + z)^4 + \Omega_M (1 + z)^3 + \Omega_X g(z)
\label{eq: hube}
\end{equation}
with $E(z) = H(z)/H_0$, $\Omega_i = \rho_i(z = 0)/\rho_{crit}$ the today density parameter for the $i$\,-\,th component, the subscripts $r, M, X$ denoting quantities referring to radiation, matter (which means the sum of baryons and dark matter) and inflessence respectively, and we define $g(z) \equiv \rho_X(z)/\rho_X(z = 0)$. From Eq.(\ref{eq: rhovsz}), we easily get\,:

\begin{equation}
g(z) = (1 + z)^{3 - \gamma} 
\left ( \frac{2 + z + z_I}{2 + z_I} \right )^{\beta} 
\left ( \frac{2 + z + z_Q}{2 + z_Q} \right )^{\gamma} \ .
\label{eq: gvsz}
\end{equation}
Note that, since we are working in a spatially flat universe, we may express $\Omega_M$ in terms of the remaining two density parameters.  

\begin{figure}
\centering \resizebox{8.5cm}{!}{\includegraphics{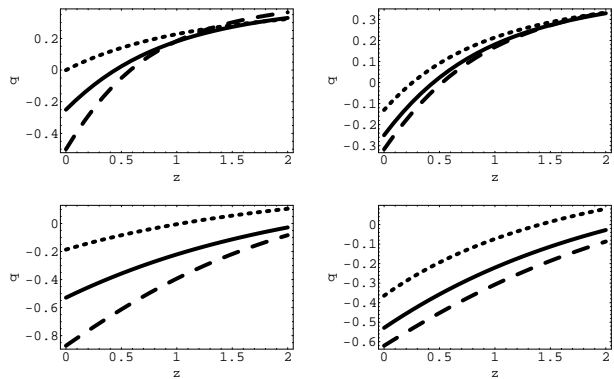}}
\caption{The deceleration parameter $q$ as function of the redshift $z$ for models with $(\beta, z_I) = (-3, 3454)$. In the left panels, $z_Q = 1.5$ is set and the three curves refer to $\gamma = 2$ (short dashed), $\gamma = 3$ (solid) and $\gamma = 4$ (long dashed). In the right panels, instead, we fix $\gamma = 3$ and consider three different values of $z_Q$, namely $z_Q = 0.5, 1.5, 2.5$ (short dashed, solid and long dashed respectively). Upper panels refer to the case $\Omega_X = 0.7$, while the lower ones are for $\Omega_X = 0.96$.}
\label{fig: qz}
\end{figure}

Let us now consider the deceleration parameter $q = - a \ddot{a}/\dot{a}^2 = -1 + (1+z) d\ln{E(z)}/dz$. Fig.\,\ref{fig: qz} shows $q(z)$ for models with $(\beta, z_I) = (-3, 3454)$ and different values of the parameters $(\gamma, z_Q, \Omega_X)$. For this latter quantity, in particular, we consider a realistic case ($\Omega_X = 0.72$) and a more extreme one ($\Omega_X = 0.96$) that allows us to investigate the consequences of neglecting the dark matter term (so that $\Omega_M = \Omega_b \simeq 0.04$, being $\Omega_b$ the baryons density parameter). 

As a general remark, we note that the higher is $\gamma$ or $z_Q$, the lower is the deceleration parameter at a given $z$, i.e. the more the universe is accelerating. As a consequence, the transition redshift $z_T$, defined by the condition $q(z_T) = 0$ and marking the transition from deceleration to acceleration, is an increasing function of both these parameters. This may be easily understood considering that $z_Q$ controls the transition from the matter\,-\,like to the dark energy\,-\,like scaling of the inflessence energy density, while $\gamma$ plays the main role in determining the value of $w_0$. Therefore, the higher is $z_Q$, the larger is the redshift range over which inflessence behaves as matter (and hence $q > 0$) so that $z_T$ increases. Similarly, the higher is $\gamma$, the higher (in absolute value) is $w_0$ so that the universe accelerates for a longer period and $z_T$ increases. As a cross check, one may study the behaviour of $z_T$ with the parameters $(\gamma, z_Q)$ which demands for a numerical investigation since the equation $q(z_T) = 0$ may not be solved analytically. Actually, such an analysis only partially confirms what we have said above. Indeed, while it is true that $z_T$ increases with $z_Q$ even if a saturation effect appears for $z_Q > 2 \div 3$, it turns out that $z_T$ may also decrease with $\gamma$ for values of $\gamma > 4 \div 5$. However, because of the interplay between $\gamma$ and $z_Q$, it is actually quite difficult to infer a general result. 

Not surprisingly, the deceleration parameter is a decreasing function of $\Omega_X$, i.e. the larger is $\Omega_X$, the more the universe accelerates. This is quite obvious since the higher is $\Omega_X$, the more the negative pressure inflessence fluid dominates the energy budget giving rise to accelerated expansion. As a consequence, $z_T$ is an increasing function of $\Omega_X$ whatever are the values of $\gamma$ and $z_Q$. It is worth noting that, since observations argue in favour of $z_T \sim 0.5$ \cite{Riess04}, the case $\Omega_X = 0.96$ is clearly disfavoured thus suggesting that inflessence may not play the role of unified dark energy in agreement with what we have inferred above on the basis of the behaviour of the EOS.

\begin{figure}
\centering \resizebox{8.5cm}{!}{\includegraphics{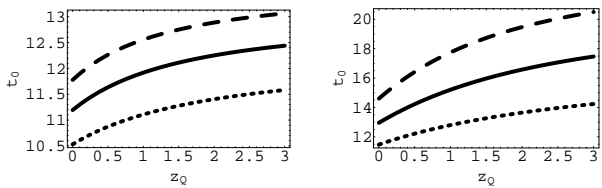}}
\caption{The age of the universe $t_0$ (in Gyr) as function of $z_Q$ for models with $(\beta, z_I, h) = (-3, 3454, 0.664)$ and three different values of $\gamma$, namely $\gamma = 2$ (short dashed), $\gamma = 3$ (solid) and $\gamma = 4$ (long dashed). Left panel refers to the realistic case $\Omega_X = 0.7$, while the right one to $\Omega_X = 0.96$.}
\label{fig: tz}
\end{figure}

As a final issue, let us determine the evolution of the scale factor $a$ with the cosmic time $t$. To this aim, we first integrate $H(z)$ to get the age of the universe at $z$\,:

\begin{equation}
t(z) = t_H \int_{z}^{\infty}{\frac{dz'}{(1 + z') E(z')}}
\label{eq: tz}
\end{equation}
where $t_H = 9.773 h^{-1} \ {\rm Gyr}$ is the Hubble time with $h$ the present day Hubble constant in units of $100 \ {\rm km \ s^{-1} \ Mpc^{-1}}$. Eq.(\ref{eq: tz}) cannot be integrated analytically in the general case, but it is straightforward to compute numerically. In particular, setting $z = 0$ as the lower limit of integration, we get the present day age of the universe $t_0$ that is shown in Fig.\,\ref{fig: tz}. From the plot, it is clear that $t_0$ is an increasing function of both $\gamma$ and $z_Q$ with the former parameter playing a more significant role. On the other hand, $t_0$ strongly depends also on $\Omega_X$ with larger values of $\Omega_X$ leading to higher values of the age of the universe. In particular, this provides a further argument against inflessence as a unified dark energy model since, for values of $\Omega_X$ close to unity, $t_0$ may become unrealistically high.

Finally, we numerically invert Eq.(\ref{eq: tz}) and use $a = (1 + z)^{-1}$ to get the dependence of the scale factor on the cosmic time. It turns out that the dependence of the scale factor on the parameters $(\gamma, z_Q)$ is more or less important depending on the value of $\Omega_X$ being quite weak for realistic values of $\Omega_X$ itself. Only if the inflessence plays the role of unified dark energy, the dependence of $a(t)$ on $(\gamma, z_Q)$ is significant. In each case, however, we note that, for a given $t$, the scale factor is larger for lower values of $\gamma$ and $z_Q$. It is worth noting that $a(t)$ does not diverge for finite $t$ even if $w_0$ is smaller than -1 for some of the combinations $(\gamma, z_Q)$ so that, as yet suggested above, the Big Rip is avoided.

\section{Matching with observations}

Whatever model which aim is to describe the evolution of the universe must be able to reproduce what is indeed observed. It is thus mandatory to test the viability of the proposed inflessence scenario by contrasting and comparing it to the astrophysical data available up to now. This is also a powerful tool to constrain the model parameters thus paving the way towards a complete characterization of the model. As an interesting byproduct, constraining the characterizing parameters allows to estimate some quantities common to every model (such as $q_0$, $t_0$, $w_0$, $z_T$) to previous values in literature. 

\subsection{The method}

In order to constrain the inflessence model parameters, we maximize the following likelihood function\,:

\begin{equation}
{\cal{L}} \propto \exp{\left [ - \frac{\chi^2({\bf p})}{2} \right ]}
\label{eq: deflike}
\end{equation}
where {\bf p} denotes the set of model parameters and the pseudo\,-\,$\chi^2$ merit function is defined as\,:

\begin{eqnarray}
\chi^2({\bf p}) & = & \sum_{i = 1}^{N}{\left [ \frac{y^{th}(z_i, {\bf p}) - y_i^{obs}}{\sigma_i} \right ]^2} \nonumber \\
~ & + & 
\displaystyle{\left [ \frac{{\cal{R}}({\bf p}) - 1.716}{0.062} \right ]^2} 
+ \displaystyle{\left [ \frac{{\cal{A}}({\bf p}) - 0.469}{0.017} \right ]^2}  \ .
\label{eq: defchi}\
\end{eqnarray}
Let us discuss briefly the different terms entering Eq.(\ref{eq: defchi}). In the first one, we consider the dimensionless coordinate distance $y$ to an object at redshift $z$ defined as\,:

\begin{equation}
y(z) = \int_{0}^{z}{\frac{dz'}{E(z')}} 
\label{eq: defy}
\end{equation}
and related to the usual luminosity distance $D_L$ as $D_L = (1 + z) y(z)$. Daly \& Djorgovki \cite{DD04} have compiled a sample comprising data on $y(z)$ for the 157 SNeIa in the Riess et al. \cite{Riess04} Gold dataset and 20 radiogalaxies from \cite{RGdata}, summarized in Tables\,1 and 2 of \cite{DD04}. As a preliminary step, they have fitted the linear Hubble law to a large set of low redshift ($z < 0.1$) SNeIa thus obtaining\,:

\begin{displaymath}
h = 0.664 \pm 0.008 \ . 
\end{displaymath}
We thus set $h = 0.664$ in order to be consistent with their work, but we have checked that varying $h$ in the $68\%$ CL quoted above does not alter the main results. Furthermore, the value we are using is consistent also with $H_0 = 72 \pm 8 \ {\rm km \ s^{-1} \ Mpc^{-1}}$ given by the HST Key project \cite{Freedman} based on the local distance ladder and with the estimates coming from the time delay in multiply imaged quasars \cite{H0lens} and the Sunyaev\,-\,Zel'dovich effect in X\,-\,ray emitting clusters \cite{H0SZ}.

The second term in Eq.(\ref{eq: defchi}) makes it possible to extend the redshift range over which $y(z)$ is probed resorting to  the distance to the last scattering surface. Actually, what can be determined from the CMBR anisotropy spectrum is the so called {\it shift parameter} defined as \cite{WM04,WT04}\,:

\begin{equation}
R \equiv \sqrt{\Omega_M} y(z_{ls})
\label{eq: defshift}
\end{equation}
where $z_{ls}$ is the redshift of the last scattering surface which can be approximated as \cite{HS96},:

\begin{equation}
z_{ls} = 1048 \left ( 1 + 0.00124 \omega_b^{-0.738} \right ) 
\left ( 1 + g_1 \omega_M^{g_2} \right )
\label{eq: zls}
\end{equation}
with $\omega_i = \Omega_i h^2$ and $(g_1, g_2)$ given in Ref.\,\cite{HS96}. The parameter $\omega_b$ is well constrained by the baryogenesis calculations contrasted to the observed abundances of primordial elements. Using this method, Kirkman et al. \cite{Kirk} have determined\,:

\begin{displaymath}
\omega_b = 0.0214 \pm 0.0020 \ .
\end{displaymath}
Neglecting the small error, we thus set $\omega_b = 0.0214$ and use this value to determine $z_{ls}$. It is worth noting, however, that the exact value of $z_{ls}$ has a negligible impact on the results and setting $z_{ls} = 1100$ does not change none of the constraints on the other model parameters.

Finally, the third term in the definiton of $\chi^2$ takes into account the recent measurements of the {\it acoustic peak} in the large scale correlation function at $100 \ h^{-1} \ {\rm Mpc}$ separation detected by Eisenstein et al. \cite{Eis05} using a sample of 46748 luminous red galaxies (LRG) selected from the SDSS Main Sample \cite{SDSSMain}. Actually, rather than the position of acoustic peak itself, a closely related quantity is better constrained from these data defined as \cite{Eis05}\,:

\begin{equation}
{\cal{A}} = \frac{\sqrt{\Omega_M}}{z_{LRG}} \left [ \frac{z_{LRG}}{E(z_{LRG})}
y^2(z_{LRG}) \right ]^{1/3}
\label{eq: defapar}
\end{equation}
with $z_{LRG} = 0.35$ the effective redshift of the LRG sample. As it is clear, the ${\cal{A}}$ parameter depends not only on the dimensionless coordinate distance (and thus on the integrated expansion rate), but also on $\Omega_M$ and $E(z)$ explicitly which removes some of the degeneracies intrinsic in distance fitting methods. Therefore, it is particularly interesting to include ${\cal{A}}$ as a further constraint on the model parameters using its measured value \cite{Eis05}\,:

\begin{displaymath}
{\cal{A}} = 0.469 \pm 0.017 \ .
\end{displaymath}
Note that, although similar to the usual reduced $\chi^2$ introduced in statistics, the reduced $\chi^2$ (i.e., the ratio between the $\chi^2$ and the number of degrees of freedom) is not forced to be 1 for the best fit model because of the presence of the priors on ${\cal{R}}$ and ${\cal{A}}$ and since the uncertainties $\sigma_i$ are not Gaussian distributed, but take care of both statistical errors and systematic uncertainties. With the definition (\ref{eq: deflike}) of the likelihood function, the best fit model parameters are those that maximize ${\cal{L}}({\bf p})$. However, to constrain a given parameter $p_i$, one resorts to the marginalized likelihood function defined as\,:

\begin{equation}
{\cal{L}}_{p_i}(p_i) \propto \int{dp_1 \ldots \int{dp_{i - 1} \int{dp_{i + 1} ... \int{dp_n {\cal{L}}({\bf p})}}}}
\label{eq: defmarglike}
\end{equation}
that is normalized at unity at maximum. Denoting with $\chi_0^2$ is the value of the $\chi^2$ for the best fit model, the $1 \sigma$ confidence regions are determined by $\Delta \chi^2 = \chi^2 - \chi_0^2 = 1$, while the condition $\Delta \chi^2  = 4$ delimited the $2 \sigma$ confidence regions.

\subsection{Results}

\begin{table*}
\begin{center}
\begin{tabular}{|c|c|c|c|c|c|c|c|}
\hline
$~$ & $\gamma$ & $z_Q$ & $\Omega_X$ & $w_0$ & $q_0$ & $z_T$ & $t_0$ (Gyr)\\
\hline \hline
{\it bf} & 3.73 & 0.1 & 0.717 & -1.03 & -0.53 & 0.65 & 14.3 \\
$1 \sigma$ & (3.42, 4.24) & $\le 1.9$ & (0.698, 0.736) & (-1.16, -0.89) & (-0.75, -0.43) & (0.58, 0.70) & (13.75, 14.86) \\
$2 \sigma$ & (3.17, 5.86) & $ \le 5.3$ & (0.678, 0.754) & (-1.52, -0.65) & (-1.15, -0.19) & (0.50, 0.75) & (13.05, 15.52) \\
\hline
\end{tabular}
\end{center}
\caption{Summary of the results of the likelihood analysis for the inflessence model. The maximum likelihood value ($bf$) and $1 \sigma$ and $2 \sigma$ range are reported for the model parameters $(\gamma, z_Q, \Omega_X)$ and for the derived quantities $(w_0, q_0, z_T, t_0)$. See text for details. Note that we are able to give only upper limits on $z_Q$ whose best fit is at the lower end of the range explored.}
\end{table*}

Before analyzing the results, it is mandatory to assess what is the parameter space to explore. In principle, when fitting the inflessence model to the observational constraints discussed above, we should assign the values of eight parameters, namely the slope parameters $(\beta, \gamma)$, the scaling redshifts $(z_I, z_Q)$, the present day density parameters of the inflessence, radiation and baryons $(\Omega_X, \Omega_r, \Omega_b)$ and the Hubble constant $h$. Given such a large parameter space, it is quite unlikely that meaningful constraints could be derived. Actually, we are interested only in a subspace of this 8D space. Indeeds, we have already set $\omega_b = \Omega_b h^2$ as explained above, while $h$ has been fixed to the same value of Ref.\,\cite{DD04} in order to be consistent with their estimate of $y^{obs}(z)$ for radio galaxies. We also set $\Omega_r = 9.89 \times 10^{-5}$ \cite{DorLil}. We may further reduce the parameter space noting that, since $z_I$ is very high, both the EOS and the expansion rate (which are the quantities probed by the observational data) are independent of $\beta$ and $z_I$. As a consequence, both these parameters are unconstrained and may be set {\it a priori} without affecting anyway the results. Motivated by these considerations, we fix $\beta = -3$ (so that the EOS asymptotically approaches the cosmological constant value) and $z_I = 3454$, i.e. the redshift of radiation\,-\,matter equivalence\footnote{Of course, this does not mean that inflation takes place soon before $z = 3454$. Indeed, the inflessence EOS approaches the value $w = -1$ only for $z >> z_I$. It is thus likely that a more correct value should be of order $10^{4 \div 6}$, but whatever choice is equivalent from the point of view of the effect on the constraints on the other parameters.}. 

\begin{figure}
\centering \resizebox{8.5cm}{!}{\includegraphics{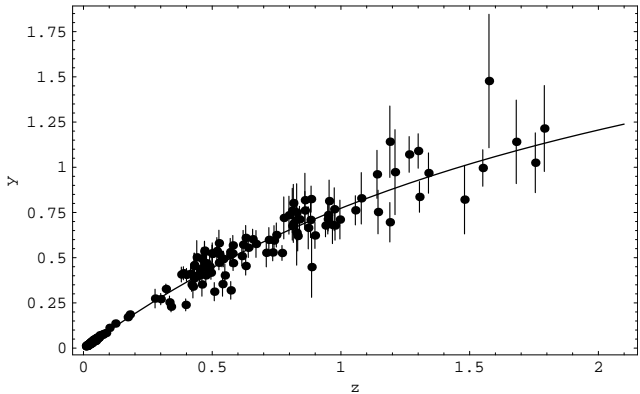}}
\caption{Observed and predicted dimensionless coordinate distance $y(z)$ diagram for the best fit inflessence model.}
\label{fig: bestfit}
\end{figure}

We are thus left with a three parameter space to explore defined by the slope $\gamma$, the scaling redshift $z_Q$ and the density parameter $\Omega_X$. Applying the procedure described in the previous subsection in this 3D space, we finally end with the constraints summarized in Table I, while Fig.\,\ref{fig: bestfit} shows the very good agreement between the data and the theoretical $y(z)$ for the best fit model. 

\begin{figure}
\centering \resizebox{8.5cm}{!}{\includegraphics{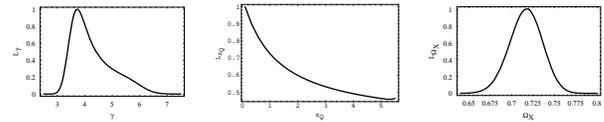}}
\caption{Marginalized likelihood functions (normalized to be 1 at maximum) for the inflessence model parameters $\gamma$ (left panel), $z_Q$ (central panel) and $\Omega_X$ (right panel).}
\label{fig: like}
\end{figure}

It is worth noting that the likelihood analysis makes it possible to efficiently constraints both $\gamma$ and $\Omega_X$ (and hence $\Omega_M$), but only upper limits may be given for $z_Q$. This is also clear looking at the marginalized likelihood functions reported in Fig.\,\ref{fig: like}. This could be qualitatively explained considering that $z_Q$ only controls when the dark energy\,-\,like behaviour starts dominating. However, we only need that $\Omega_X$ is larger than $\Omega_M$ and the EOS is negative today in order to be able to have an accelerated expansion in the redshift range probed by the SNeIa and radio galaxies. As such, it is not just necessary that $z_Q > 0$ to fit the data thus explaining why the constraints on this parameter are so weak.   

In Table I, we also give the estimated values of some physically interesting quantities, namely the present day values of the EOS $w_0$ and the deceleration parameter $q_0$, the transition redshift $z_T$ defined by the condition $q(z_T) = 0$ and the age of the universe $t_0$. Since the uncertainties on the model parameters are not Gaussian distributed, a naive propagation of the errors is not possible. We thus estimate the 1 and $2 \sigma$ constraints on the derived quantities by randomly generating 20000 points $(\gamma, z_Q, \Omega_X)$ using the marginalized likelihood functions of each parameter and then constructing the likelihood function of the derived quantity. Although not statistically well motivated, this procedure gives a conservative estimate of the uncertainties.

It is quite interesting to compare our constraints on some derived quantities to those previously available in literature. First, we consider the total matter density parameter $\Omega_M$ that could be derived from our estimate of $\Omega_X$ thus obtaining $\Omega_M = 0.283 \pm 0.019$ (at $1 \sigma$). This is in very good agreement with other estimates such as $\Omega_M = 0.29^{+0.05}_{-0.03}$ obtained by Riess et al. \cite{Riess04} by fitting the quiessence model (i.e. dark energy with constant equation of state) and $\Omega_M = 0.284_{-0.060}^{+0.079}$ (at $99\%$ CL) found by Seljak et al. \cite{Sel04} through a combined analysis of the CMBR anisotropy spectrum measured by WMAP, the power spectrum of SDSS galaxies, the SNeIa Gold dataset, the dependence of the bias on luminosity and the Ly$\alpha$ power spectrum in the framewrok of the concordance $\Lambda$CDM model. It is also consistent with the Allen et al. \cite{A04} result $\Omega_M = 0.245^{+0.040}_{-0.037}$ from the fitting to the gas mass fraction in galaxy cluysters. Overall, there is a very good agreement with our result which is encouraging since, whatever is the dark energy component, the matter abundance should be the same. 

As a second check, we consider the age of the universe $t_0$ whose best fit value turns out to be 14.3 Gyr in good agreement with $t_0 = 13.24_{+0.41}^{+0.89} \ {\rm Gyr}$ from Tegmark et al. \cite{Teg03}, $t_0 = 14.4^{+1.4}_{-1.3}\ {\rm Gyr}$ found by Rebolo et al. \cite{VSA} and $t_0 = 13.6 {\pm} 01.19 \ {\rm Gyr}$ given by Seljak et al. \cite{Sel04}. Although such results are rigorously model dependent, the agreement among the different estimates and our own leads furhter support to the inflessence scenario.

It is more interesting to consider the constraints on the transition redshift $z_T$. It is difficult to derive this quantity directly from the data even if some attempts have been made. Fitting the phenomenological parametrization $q(z) = q_0 + dq/dz|_{z = 0} z$ to the SNeIa Hubble diagram, Riess et al. \cite{Riess04} have found $z_T = 0.46 {\pm} 0.13$ (at $1 \sigma$ level) which is in agreement with our result only at the $2 \sigma$ level. Since it is not clear what is the systematic error induced by the linear approximation of $q(z)$, which only works over a limited redshift range, we do not consider a serious shortcoming of the model the $1 \sigma$ disagreement between this estimate of $z_T$ and our one.

As a final remark, it is intriguing to note that, although the 1 and $2 \sigma$ ranges extend in the phantom like region, the best fit value of $w_0$ is only marginally different from that of the cosmological constant. Considering also the estimated $\Omega_X$, we may conclude that the best fit inflessence model is today fully equivalent to the concordance $\Lambda$CDM model which explains why we are able to fit equally well the observational data.   

\subsection{The lookback time to galaxy clusters}

In a recent paper \cite{FA}, some of us have proposed to use the lookback time to galaxy clusters and the age of the universe as an alternative test to investigate the viability of a given cosmological model and to constrain its parameters. Being based on a different astrophysics (stellar evolutionary processes rather than calibrating standard candles) and on a different physical quantity (time rather than distance), this method is a valid alternative to the usual cosmological tests. Since, unfortunately, at the moment the dataset is very limited, the method may be considered as a cross check of the results obtained by the usual probes and in this sense we will apply it to the inflessence model.

Here, we briefly report the basics of the method referring the interested reader to Ref.\,\cite{FA} for further details. First, let us remind that, given an object at redshift $z$, the lookback time is defined as the difference between the present day age of the universe and the age of the universe at that redshift\,:

\begin{equation}
t_L(z, {\bf p}) = t_H \int_{0}^{z}{\frac{dz'}{(1 + z') E(z')}} \ .
\label{eq: deftl}
\end{equation}
Having estimated somewhat the age of the object $t_{age}$, $t_L$ may be determined from observations as $t_L^{obs}(z) = t_0^{obs} - t_{age}(z) - df$, with $t_0^{obs}$ the estimated age of the universe and $df$ the so called {\it delay factor} which takes into account our ignorance of the formation redshift of the object (and hence of the difference between the age of the universe and that of the object). Constraints on model parameters are then obtained by a likelihood analysis similar to the one performed in the previous section, but redefining the pseudo\,-\,$\chi^2$ merit function as\footnote{Note that in \cite{FA} there is a prior on $h$ in the definition of $\chi_{lt}^2$ that we have removed here since $h$ has been set above.}\,:

\begin{equation}
\chi_{lt}^2 = \sum_{i}^{N_{lt}}{\left [ 
\frac{t_L^{th}(z_i, {\bf p}) - t_L^{obs}(z_i)}{\sqrt{\sigma_i^2 + \sigma_t^2}} \right ]^2}
+ \left [ \frac{t_0^{th}({\bf p}) - t_0^{obs}}{\sigma_t} \right ]^2
\label{eq: defchilt}
\end{equation}
with $\sigma_i$ and $\sigma_t$ the uncertainties on the age of the $i$\,-\,th cluster and the present day age of the universe respectively. Following \cite{FA}, we set $\sigma_i = 1 \ {\rm Gyr}$ although this is a gross overestimate which does not affect, however, the results. We also fix $(t_0^{obs}, \sigma_t) = (14.4, 1.4) \ {\rm Gyr}$ in agreement with \cite{VSA}.  

Having only seven data (six clusters and the age of the universe), it is almost obvious that we are not able to constrain the full parameter space $(\gamma, z_Q, \Omega_X, df)$ so that we have decided to set $\Omega_X$ to the best fit value in Table I to finally get the following constraints on $(\gamma, z_Q)$\,:

\begin{equation}
\begin{array}{ll}
3.2 \le \gamma \le 5.4 \ , \ z_Q \le 5.1 \ , \ 3.1 \le df \le 4.6 & {\rm at \ 1 \sigma} \nonumber \\
~ & ~ \\ 
2.5 \le \gamma \le 7.3 \ , z_Q \le \infty \ , \ 2.3 \le df \le 5.3 & {\rm at \ 2 \sigma} \nonumber \\
\end{array}
\end{equation}
with $(\gamma, z_Q, df) = (4.1, 0.1, 3.8)$ as best fit parameters and $df$ in Gyr. Not surprisingly, the constraints on $(\gamma, z_Q)$ are significantly weakened with respect to the results in Table I so that at $2 \sigma$ level the upper limit on $z_Q$ is formally unconstrained (i.e., it is larger than the upper end of the range probed).  However, it is worth noting that the constraints on $\gamma$ are in very good agreement with the ones obtained before and also the best fit value is quite close. Moreover, the delay factor turns out to be quite similar to that estimated using radically different models (see \cite{FA} for details) which is a reassuring result since this quantity is mainly determined by astrophysical processes and should be hence model independent. We may therefore conclude that this test leads further support to the result obtained by matching the inflessence model to the observational data discussed above.

\section{Theoretical motivations}

The key ingredient of the inflessence model we are investigating is the scaling of energy density with the scale factor given by Eq.(\ref{eq: rhovsa}). This ansatz has been phenomenologically motivated by the need to have a fluid which is able to account for both inflation and the present day accelerated expansion with the further advantage of tracking the matter for a long period thus alleviating the coincidence problem. The very good matching with two independent datasets is an {\it a posteriori} justification of the model and strongly motivates further studies. Nevertheless, Eq.(\ref{eq: rhovsa}) is not based on any underlying physical theory. Actually, given its phenomenological nature, finding a unique theoretical background for the inflessence scenario is not easy. Here, we propose two possible and radically different mechanisms which are able to give rise to a scenario with a single effective fluid having the same properties of the inflessence. 

\subsection{Scalar field quintessence}

Since inflessence is able to fit the observational data with the same accuracy as other popular dark energy candidates, it is worth wondering whether it is possible to recover Eq.(\ref{eq: rhovsa}) resorting to the same kind of mechanism. To this end, it is useful to remind that, in most of the quintessence models, the cosmic acceleration is driven by a scalar field $\phi$ rolling down its self interaction potential and giving rise to a fluid whose energy density and pressure read\,:

\begin{equation}
\left \{
\begin{array}{l}
\rho = \displaystyle{\frac{1}{2} \dot{\phi}^2 + V_{\phi}(\phi)} \ , \nonumber \\
~ \\
p = \displaystyle{\frac{1}{2} \dot{\phi}^2 - V_{\phi}(\phi)} \ , \nonumber \\
\end{array}
\right .
\label{eq: rpsc}
\end{equation}
so that the EOS is\,:

\begin{equation}
w_{\phi} = p/\rho = \frac{1 - 2 V_{\phi}(\phi)/\dot{\phi}^2}{1 + 2 V_{\phi}(\phi)/\dot{\phi}^2} \ .
\label{eq: wphi}
\end{equation}
It is easy to see that $w_{\phi} < -1$ is not allowed by Eq.(\ref{eq: wphi}), while this is possible for some combination of the inflessence model parameters $\gamma$ and $z_Q$ (and, indeed, the estimated $w_0$ in Table I does not exclude such values). Changing the sign of the kinetic term in Eqs.(\ref{eq: rpsc}), we get phantom models whose EOS is, on the contrary, always smaller than -1. The inflessence EOS may cross the so called phantom divide, i.e. the EOS may pass from a value $w < -1$ to $w > -1$ or vice versa for some combination of the model parameters (see, for instance, the long dashed curve in the left panel of Fig.\,\ref{fig: wz}). 

\begin{figure}
\centering \resizebox{8.5cm}{!}{\includegraphics{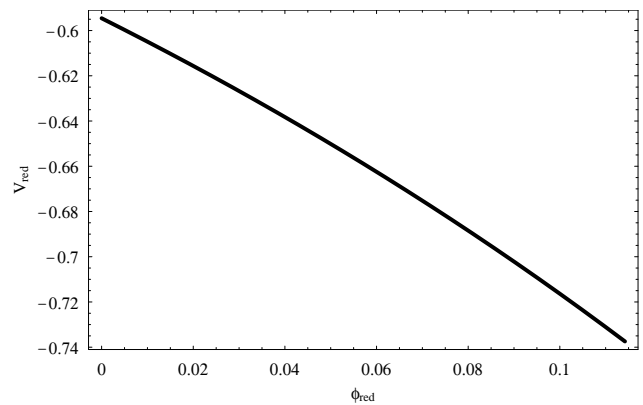}}
\caption{Reconstructed scalar field potential for the inflessence model with parameter values as detailed in the text. Note that, for numerical convenience, we plot $V_{red} = V/\rho_{crit}$ vs $\phi_{red} = H_0 \phi/\rho_{crit}$ so that the numerical values are somewhat arbitrary. The reconstruction shown refers to the redshift range $(0, 3)$, but nothing prevents to extend it to higher $z$.}
\label{fig: vpotfig}
\end{figure}

Actually, since $z_I$ is very large, for $z < z_I$, Eq.(\ref{eq: wvsz}) may be very well approximated as\,:
\begin{equation}
w \simeq - \frac{\gamma}{3} \left ( \frac{1 + z_Q}{2 + z + z_Q} \right )
\end{equation}
and it is easy to show that $w \ge -1$ for $z > z_{\Lambda} = - 1 + (1 + z_Q)(1 + \gamma/3)$. In order to have acceleration, $w_0 \le 1/3$ is needed. Imposing this condition, we get $z_{\Lambda} < 0$ so that the approximated EOS is always larger than -1 and we may recover the inflessence model by resorting to a single scalar field. From Eqs.(\ref{eq: rpsc}), we get\,:

\begin{equation}
\left \{
\begin{array}{l}
\phi(z) = \displaystyle{\frac{\sqrt{\Omega_X \rho_{crit}}}{H_0} 
\int_{0}^{z}{\frac{\sqrt{\left [ 1 + w(z') \right ] g(z')}}{(1 + z') E(z')}}} \nonumber \\
~ \\
V(z) = \displaystyle{-\frac{\Omega_X \rho_{crit}}{2} \left [ 1 - w(z) \right ] g(z)} \nonumber
\end{array}
\right .
\label{eq: vphi}
\end{equation}
with $w(z)$ and $g(z)$ given by Eqs.(\ref{eq: wvsz}) and (\ref{eq: gvsz}) respectively. Eliminating numerically $z$ from this set of equations, we finally reconstruct the potential $V(\phi)$ which is shown in Fig.\,\ref{fig: vpotfig} for the inflessence model with $(\beta, \gamma, z_Q, z_I, \Omega_X, h) = (-3, 3.73, 0.1, 3454, 0.72, 0.664)$. As it is clear from the plot, the potential may be very well approximated by a simple power law with negative slope. Such a shape of the potential is quite common in quintessential scenarios.

For completeness, we remind that, for those combinations of parameters that give rise to inflessence EOS crossing the phantom divide, the interpretation in terms of a single scalar field must be abandoned. However, it is possible to recover again Eq.(\ref{eq: rhovsa}) resorting to the {\it quintom} scenario \cite{quintom} where both a scalar and a phantom field are introduced with an eventual interaction term. In an alternative scenario of the same kind, dubbed {\it hessence}, the role of the two fields is played by the real and the imaginary part of a single complex scalar field \cite{hessence}. Both quintom and hessence offer the possibility of recovering the inflessence model for those combination of $(\beta, \gamma, z_Q, z_I, \Omega_X)$ leading to $w(z)$ crossing the phantom divide.

\subsection{Inflessence and $f(R)$ gravity}

\begin{figure}
\centering \resizebox{8.5cm}{!}{\includegraphics{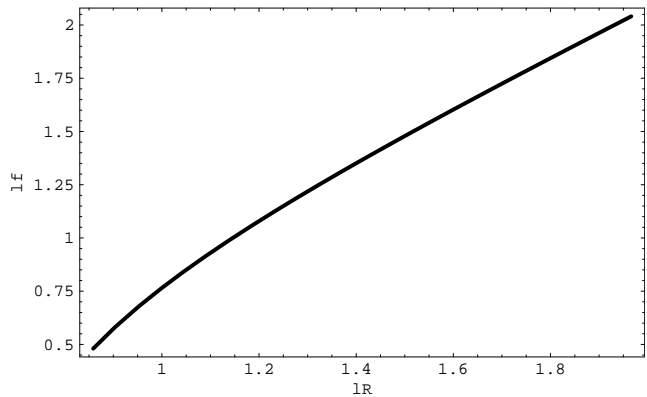}}
\caption{Reconstructed $f(R)$ for the inflessence model with parameters set as in Fig.\,\ref{fig: vpotfig}. Note that we plot $lf = \log{(-f)}$ vs $lR = \log{(-R)}$ for numerical convenience and perform the reconstruction over the redshift range (0, 3).}
\label{fig: frfig}
\end{figure}

Recently, it has been demonstrated \cite{CCT} that, under some quite general conditions, the dynamics (i.e., the scale factor and the expansion rate) of dark energy model may be obtained as the byproduct of a fourth order theory of gravity. In such a framework, standard matter (both dust and radiation) is the only fluid filling the universe, but the correct theory of gravity is no more the Einsteinian general relativity, but a higher order theory obtained by replacing the Ricci scalar $R$ in the gravity Lagrangian with an analytic function $f(R)$. It is then possible \cite{noicurv} to show that the dynamical equations may still be written in the usual form provided that $(i)$ the matter term is coupled to the geometry through $f'(R)$ with the prime denoting derivative with respect to $R$ and $(ii)$ an effective {\it curvature} fluid is defined with energy density and EOS determined by $f(R)$ and its derivatives up to the third order\footnote{Because of the presence of this fluid, such scenario is also referred to as {\it curvature quintessence}.}. 
 
The method developed in \cite{CCT} makes it possible to derive the $f(R)$ theory which reproduces a given $H(z)$ and may thus be straightforwardly applied to the case of the inflessence model. We do not discuss the basics of the method referring the interested reader to the quoted paper, while we show the result in Fig.\,\ref{fig: frfig}. Although an analytical solution is not available, we may approximate the reconstructed $f(R)$ very well (with an error less than $1\%$) as\,:

\begin{equation}
f(R) \simeq - {\rm dex}(f_0 + f_1 lR + f_2 lR^2 + f_3 lR^3)
\label{eq: frapprox}
\end{equation}
where $dex(x) = 10^x$, $lR = \log{(-R)}$ and $(f_0, f_1, f_2, f_3)$ depend on the inflessence model parameters. In particular, for the choice in figure, we find\,:

\begin{displaymath}
(f_0, f_1, f_2, f_3) = (-2.15, 4.45, -1.92, 0.377) \ .
\end{displaymath} 
We stress, however, that Eq.(\ref{eq: frapprox}) is only a fitting function working well over the redshift range considered in the reconstruction and it has not to be meant as the actual $f(R)$ to be extrapolated to higher $z$. 

\section{Conclusions}

The discovery of the present day cosmic acceleration of the universe has opened the way to a wide and still ongoing debate on what is driving such an accelerated expansion. Moreover, the evidences in favour of a spatially flat and low matter universe coming from the CMBR anisotropy spectrum and the large scale structure, respectively, has made this problem more intriguing.

Rather than adding a further somewhat theoretically motivated proposal to the large list yet available, we have here pursued a different approach to the problem. To this end, we have first noted that cosmic acceleration is not a unique event since the universe has already undergone a phase of accelerated expansion in its very early history, i.e. during the inflationary epoch. It is clearly desirable to have a mechanism able to produce an initial period of accelerated expansion followed by a long phase of matter domination (not interfering with nucleosynthesis and structure formation) finally ending up with the present day cosmic acceleration. Motivated by these purely phenomenological considerations, we have introduced in the energy budget a fluid whose energy density is postulated to be given by Eq.(\ref{eq: rhovsa}). By suitably choosing the model parameters, the EOS approaches the cosmological constant value $(w_{\Lambda} = -1)$ both for $z \rightarrow \infty$ and for $z \rightarrow -\infty$ so that a de Sitter state is recovered in the very early and very late universe. Moreover, the fluid energy density scales with the scale factor $a$ in the same way as ordinary (dust) matter for most of the universe history thus solving (or, at least, alleviating) the coincidence problem. Given its ability to give rise to both inflationary and quintessential behaviour, we have dubbed {\it inflessence} such a fluid. 

A further support to the inflessence scenario is given by the matching with observations. Considering the dimensionless coordinate distance to Gold SNeIa sample and a dataset comprising 20 radio galaxies, the shift parameter and the baryonic acoustic peak in the LRG correlation function, we have been able not only to show the viability of the model, but also to constrain its main parameters. The constraints derived on some other interesting physical quantities (such as the present day values of the EOS, the deceleration parameter, the transition redshift and the age of the universe) are in very good agreement with previous ones in literature which also comprise some model independent estimates.  As an independent cross check, we have also used a recently proposed method to constrain the model parameters with the lookback time to galaxy clusters and the age of the universe thus obtaining consistent estimates for the model parameters. 

These results further support the inflessence scenario and represent an {\it a posteriori} confirmation of the phenomenological ansatz we are dealing. Nevertheless, we have also wondered whether it is possible to recover Eqs.(\ref{eq: rhovsz}) and (\ref{eq: wvsz}) by resorting to some theoretically motivated mechanism. Two approaches have been investigated. In the first case, inflessence turns out to be an effective fluid resulting from the combination of a scalar and a phantom field evolving under their self interaction potentials with a possible interaction between them. However, for the best fit model parameters, a single scalar field is necessary and we have numerically reconstructed both the evolution of the field and its potential. A radically different interpretation of inflessence is connected with higher order gravity theories. In such a case, the same dynamics generated by the matter and inflessence contributions to the energy density is recovered in the framework of fourth order theories of gravity. In particular, using a recently derived formalism, we have been able to numerically determine the function $f(R)$ entering the generalized gravity Lagrangian. It is worth noting that such an approach allows to find out a particular class of extended theories of gravity able to explain both inflation and the present day cosmic acceleration. 

The successful results obtained are encouraging and motivate further studies of the inflessence scenario. From an observational point of view, it is interesting to compare and contrast the model with the CMBR anisotropy and polarization spectrum and with the data on the matter power spectrum and the growth index. These tests make it possible to check the model over a different redshift range than the SNeIa and radio galaxies data offering also the possibility to tighten the ranges for the different parameters (in particular $z_Q$). It should also be desirable to better investigate the theoretical background of inflessence. Although two possible mechanisms have been proposed, there is still the possibility that other approaches may lead to Eq.(\ref{eq: rhovsa}) either resorting to scalar fields or some extended theories of gravity and modified Friedman equations (e.g., connected with branes). To this regard, it is worth stressing that such theoretical model should be able to explain at the same time two of the most exciting problems of modern cosmology, namely inflation and present day cosmic acceleration. As a natural consequence, the coincidence problem results to be greatly alleviated as it is typical of all unified models.

\end{document}